\documentclass[12pt,epsf]{article}
\usepackage{graphicx}
\usepackage{amssymb,amsmath}

\begin{document}

\def\a{\alpha}
\def\b{\beta}
\def\c{\varepsilon}
\def\d{\delta}
\def\e{\epsilon}
\def\f{\phi}
\def\g{\gamma}
\def\h{\theta}
\def\k{\kappa}
\def\l{\lambda}
\def\m{\mu}
\def\n{\nu}
\def\p{\psi}
\def\q{\partial}
\def\r{\rho}
\def\s{\sigma}
\def\t{\tau}
\def\u{\upsilon}
\def\v{\varphi}
\def\w{\omega}
\def\x{\xi}
\def\y{\eta}
\def\z{\zeta}
\def\D{\Delta}
\def\G{\Gamma}
\def\H{\Theta}
\def\L{\Lambda}
\def\F{\Phi}
\def\P{\Psi}
\def\S{\Sigma}
\def\BR{{\rm Br}}
\def\o{\over}
\def\beq{\begin{eqnarray}}
\def\eeq{\end{eqnarray}}
\newcommand{\nn}{\nonumber \\}
\newcommand{\gsim}{ \mathop{}_{\textstyle \sim}^{\textstyle >} }
\newcommand{\lsim}{ \mathop{}_{\textstyle \sim}^{\textstyle <} }
\newcommand{\vev}[1]{ \left\langle {#1} \right\rangle }
\newcommand{\bra}[1]{ \langle {#1} | }
\newcommand{\ket}[1]{ | {#1} \rangle }
\newcommand{\EV}{ {\rm eV} }
\newcommand{\KEV}{ {\rm keV} }
\newcommand{\MEV}{ {\rm MeV} }
\newcommand{\GEV}{ {\rm GeV} }
\newcommand{\TEV}{ {\rm TeV} }
\def\diag{\mathop{\rm diag}\nolimits}
\def\Spin{\mathop{\rm Spin}}
\def\SO{\mathop{\rm SO}}
\def\O{\mathop{\rm O}}
\def\SU{\mathop{\rm SU}}
\def\U{\mathop{\rm U}}
\def\Sp{\mathop{\rm Sp}}
\def\SL{\mathop{\rm SL}}
\def\tr{\mathop{\rm tr}}

\newcommand{\bear}{\begin{array}}  
\newcommand {\eear}{\end{array}}
\newcommand{\bea}{\begin{eqnarray}}  
\newcommand {\eea}{\end{eqnarray}}
\newcommand{\la}{\left\langle}  
\newcommand{\ra}{\right\rangle}
\newcommand{\non}{\nonumber}  
\newcommand{\ds}{\displaystyle}
\newcommand{\red}{\textcolor{red}}
\def\ubl{U(1)$_{\rm B-L}$}
\def\REF#1{(\ref{#1})}
\def\lrf#1#2{ \left(\frac{#1}{#2}\right)}
\def\lrfp#1#2#3{ \left(\frac{#1}{#2} \right)^{#3}}
\def\OG#1{ {\cal O}(#1){\rm\,GeV}}

\def\TODO#1{ {\bf ($\clubsuit$ #1 $\clubsuit$)} }


\baselineskip 0.7cm

\begin{titlepage}

\begin{flushright}
IPMU-13-0148 \\
\end{flushright}

\vskip 1.35cm
\begin{center} 
{\large \bf 
Bino-Higgsino Mixed Dark Matter in a \\
Focus Point Gaugino Mediation
} \\

\vskip 1.2cm

{ Tsutomu T. Yanagida and Norimi Yokozaki}

\vskip 0.4cm

{\it Kavli Institute for the Physics and Mathematics of the Universe (WPI),\\
Todai Institutes for Advanced Study, University of Tokyo,\\
Kashiwa 277-8583, Japan\\
}

\vskip 1.5cm

\abstract{
We investigate the neutralino dark matter in the focus point gaugino mediation model with the $\mathcal{O}(100)$ GeV gravitino. The thermal relic abundance of the neutralino with a sizable Higgsino fraction can explain the dark matter density at the present universe. The spin-independent cross section is marginally consistent with the current upper limit from the XENON 100 experiment, and the whole parameter region can  be covered at the XENON1T experiment. We also discuss the origin of the gluino mass to wino mass ratio at around 3/8, which is crucial for the mild fine-tuning in the electroweak symmetry breaking sector. It is shown that the existence of the non-anomalous discrete R-symmetry can fix this ratio to 3/8. 
 }
\end{center}
\end{titlepage}

\setcounter{page}{2}

\section{Introduction}

A gaugino medition model \cite{MYY} was proposed motivated by Linde's adiabatic solution to the Polonyi problem \cite{Linde, NTY}. This gaugino mediation model is very attractive, since the electroweak symmetry breaking (EWSB) scale can be naturally explained 
when the ratio of gluino to wino mass is set at a certain value \cite{YY};\footnote{This was also known in a more generic gravity mediation framework \cite{Kyoto}.} the EWSB scale becomes insensitive to the gaugino mass scale, i.e., the SUSY breaking scale. This behavior is similar to that of the focus point scenario~\cite{Focus}, where the generated EWSB scale is not affected significantly by the change of the universal scalar mass $m_0$, provided that $m_0$ is much larger than gaugino masses.\footnote{The focus point behavior is also discussed in a context of high scale gauge mediation models~\cite{blummer}.}

In this gaugino mediation model the gravitino mass is assumed to be $m_{3/2} = O(10)$ GeV and hence the decay of the next lightest SUSY particle (NLSP) occurs during or after the Big Bang Nucleosynthesis (BBN) and it destroys light elements produced by the BBN. Therefore, we have introduced {\it ad hoc} $R$ parity violation such that the NLSP decays into the standard model (SM) particles before the BBN~\cite{BBN_RPV}.
The assumption of the light gravitino of mass $O(10)$ GeV is based on the adiabatic solution \cite{NTY} to the Polonyi problem; the self-couplings of the Polonyi field and couplings to gauginos are assumed to be enhanced, inducing large gaugino masses. 

However, it has been, recently, found that the Polonyi problem can be easily solved by a simple extension of the Polonyi superpotential 
\cite{HISY} and hence we can introduce the Polonyi field without causing the cosmological Polonyi problem. Therefore, we do not need to invoke the adiabatic solution and hence the gravitino mass can be taken as $O(100)$ GeV mass. In this letter we discuss this new possibility where the gravitino mass is in a region of $300-600$ GeV.

The reason why we choose the above gravitino mass region is to avoid overproduction of the gravitino in the early universe (see \cite{KMY}). The mass region of $300-600$ GeV is known to be consistent with the BBN if the reheating temperature $ T_R \lesssim10^6$ GeV \cite{KMY} which is almost the lower bound of the reheating temperature for the leptogenesis  \cite{FY,BPY}. 

We first show that we still keep the focus point parameter space where only a quite mild fine tuning $\Delta \lesssim 100$ (the definition of $\Delta$ is given later) is required for the electroweak symmetry breaking even with the gravitino mass of $300-600$ GeV. Second, we point out that the thermal relic abundance of the bino-Higgsino dark matter (DM) explains the observed DM density in the focus point region. The mass and the scattering cross section of the DM is marginally allowed by the XENON100 experiment~\cite{Xenon100}. Therefore, we can exclude or confirm the present model in near future DM detection experiments. We also comment on IceCube experiment, which gives a constraint on the spin-dependent cross section stronger than that from XENON100~\cite{IceCube}. Finally, we discuss the origin of the gluino to wino mass ratio at around 3/8, relaxing the fine-tuning in the EWSB sector.

\section{The focus point gaugino mediation with $m_{3/2}=O(100)$ GeV}
In the minimal supersymmetric standard model (MSSM), the electroweak symmetry breaking (EWSB) can be explained dynamically; the EWSB is triggered by the SUSY breaking through radiative corrections mainly from top/stop loops.
In the focus point gaugino mediation~\cite{YY}, the EWSB scale can be naturally obtained even when the stop masses are large as a few TeV and the focus point gaugino mediation is consistent with the observed Higgs boson mass of around 125 GeV\,\cite{Higgs_exp, Higgs_exp2} and non-observation of the colored SUSY particles~\cite{LHC_SUSY, LHC_SUSY2}.


The EWSB scale is determined by minimization conditions of the Higgs potential:
\bea
\frac{m_{\hat{Z}}^2}{2} &=&\frac{(m_{H_d}^2 + \frac{1}{2v_d}\frac{\partial \Delta V}{\partial v_d}  ) - (m_{H_u}^2+ \frac{1}{2v_u}\frac{\partial \Delta V}{\partial v_u})\tan^2\beta}{\tan^2\beta-1}  -\mu^2, \nn
(\tan\beta+\cot\beta)^{-1}&=&\frac{B \mu}{2|\mu|^2+m_{H_u}^2 +m_{H_d}^2  + \frac{1}{2v_d}\frac{\partial \Delta V}{\partial v_d}+ \frac{1}{2v_u}\frac{\partial \Delta V}{\partial v_u}},
\eea
where $m_{H_u}$ and $m_{H_d}$ are soft masses of the up-type Higgs $H_u$ and down-type Higgs $H_d$, respectively. The radiative corrections to the Higgs potential is denoted by $\Delta V$, and $v_u=\left<H_u^0\right>$ and $v_d=\left<H_d^0\right>$ are vacuum expectation values of up- and down-type Higgs, respectively. Here, $m_{\hat Z}$ is the electroweak scale (times gauge coupling) $m_{\hat Z}^2=(1/2)(g_Y^2+g_2^2)(v_u^2 + v_d^2)$, where $g_2$ $(g_Y)$ is the gauge coupling of $SU(2)_L$ ($U(1)_Y$). The experimental value is $m_{\hat Z} \simeq 91.2$ GeV~\cite{PDG}. Note that rather large $\tan\beta(=v_u/v_d)$ of $\mathcal{O}(10)$ is required since otherwise the Higgs boson mass becomes too small. As a result, the size of $m_{H_u}^2$ is more important than that of $m_{H_d}^2$; $m_{\hat Z}$ is approximately written as $m_{\hat Z}^2 \simeq -2 (m_{H_u}^2 + \frac{1}{2v_u}\frac{\partial \Delta V}{\partial v_u}) -2\mu^2$.

\begin{figure}[t]
\begin{center}
\includegraphics[scale=1.1]{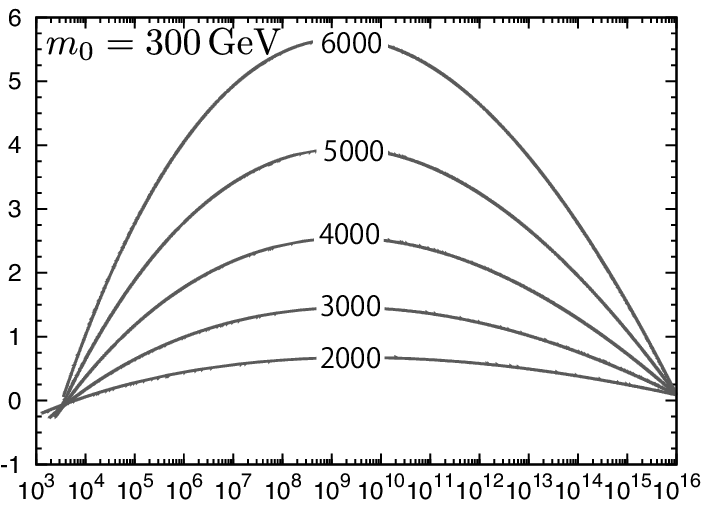}
\includegraphics[scale=1.1]{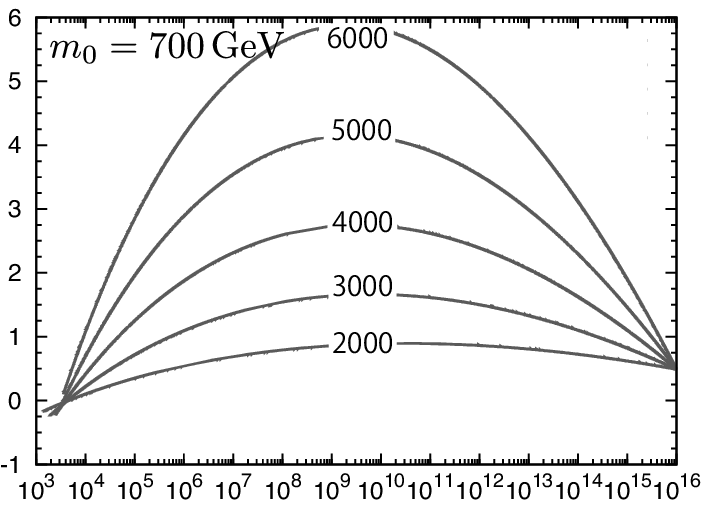}
\includegraphics[scale=1.13]{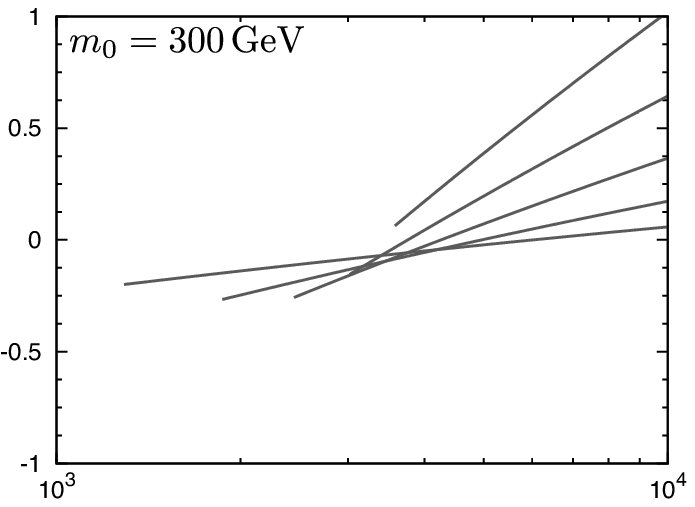}
\includegraphics[scale=1.13]{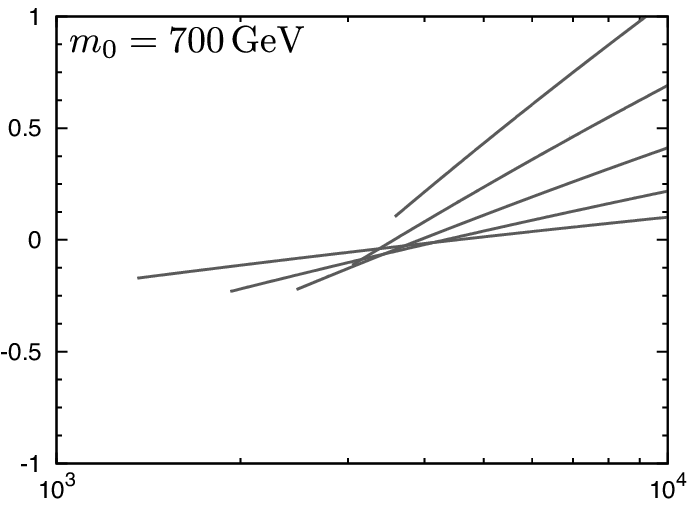}
\caption{$m_{H_u}^2$ for different $m_0$ in the unit of $({\rm TeV})^2$ as a function of the renormalization scale. The Wino masses are taken as $M_{1/2}=6, 5, 4, 3$ and $2 \,{\rm TeV}$ (from top to bottom). Here,  $\tan\beta=20$, $M_3/M_2=0.39$, $M_1/M_2=0.4$ and $\mu<0$. The SM parameters are taken as $m_t =173.2$ GeV and $\alpha_S(m_Z)=0.1184$. In the lower two panels, the regions near the weak scale are magnified.}
\label{fig:focus}
\end{center}
\end{figure}

The up-type Higgs boson mass at the soft mass scale (usually taken as the stop mass scale), can be written in terms of the parameters at the high energy scale. In our gaugino mediation model, we have the gaugino mass parameter at the GUT scale $M_{1/2}$, the universal scalar masses $m_0$ and $\tan\beta$. Here, we consider the case of $M_{1/2} = O({\rm TeV})$ while $m_0=O(100\, {\rm GeV})$. The bino, wino and gluino masses at the GUT scale are parameterized as 
\bea
(M_1, M_2, M_3)= (r_1, 1, r_3) M_{1/2}.
\eea
The ratios of the gaugino mases, $r_1$ and $r_3$, are fixed by more fundamental high energy physics.
With these GUT scale parameters, the up-type Higgs soft mass squared at the soft mass scale can be written as
\bea
m_{H_u}^2 (2.5\, {\rm TeV}) &\simeq& -1.197 M_3^2 + 0.235 M_2^2- 0.013 M_1 M_3 - 0.134 M_2 M_3 \nn
&+& 0.010 M_1^2-0.027 M_1 M_2 + 0.067 m_0^2, \label{eq:mhu2}
\eea
using two-loop renormalization group equations~\cite{2loopRGE}. Here, we take $\tan\beta=20$ and $m_t =173.2$ GeV. Notice that $m_{H_u}^2$ becomes small for $r_3 \sim 0.4$. For instance, $m_{H_u}^2(2.5\,{\rm TeV})\simeq -{\rm 0.006} M_{1/2}^2 + 0.067 m_0^2\,$  for  $r_1=r_3=(3/8)$; $m_{H_u}^2$ is not sensitive to the universal scalar mass $m_0$, since the coefficient of $m_0^2$ in Eq.(\ref{eq:mhu2}) is small and $m_0^2$ is assumed to be much smaller than $M_{1/2}^2$ in our gaugino mediation model.
In Fig.~\ref{fig:focus}, we show the focus point behavior with different $m_0$. The behavior is almost insensitive to $m_0$ as expected. Here, renormalization group equations for soft mass parameters are evaluated at the two-loop level using {\tt SuSpect} package~\cite{suspect}.

\begin{figure}[t]
\begin{center}
\includegraphics[scale=1.0]{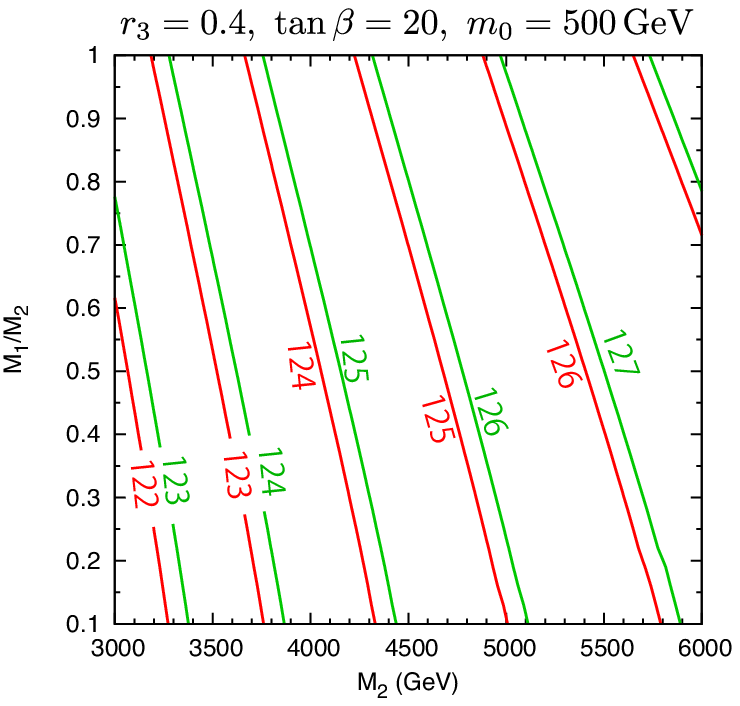}
\includegraphics[scale=1.0]{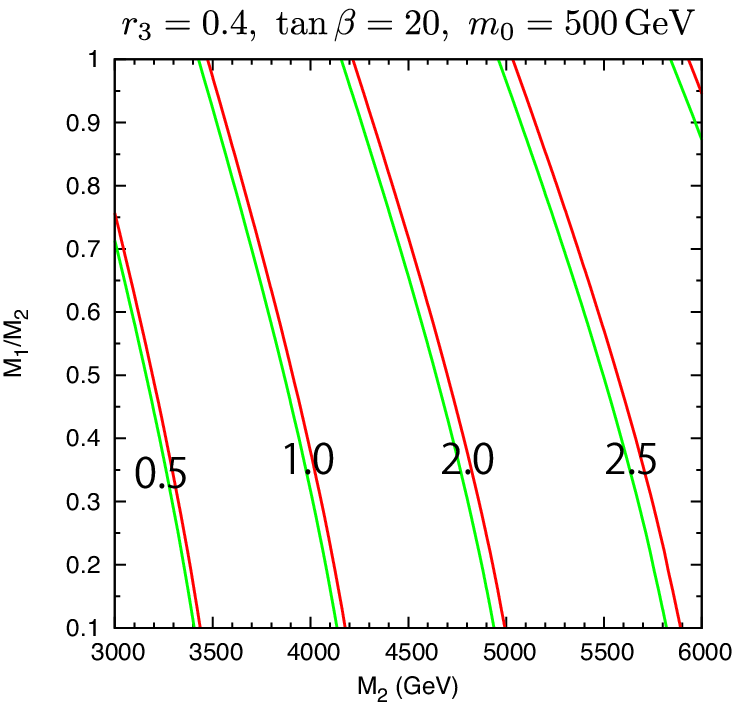}
\caption{Contours of the Higgs boson mass (left panel) and $\Delta m_h$ (right panel) in the unit of GeV. The red (green) lines drawn with the top mass of $m_t=173.2$ GeV (174.2 GeV). Here, $\alpha_S(m_Z)=0.1184$.}
\label{fig:mhiggs}
\end{center}
\end{figure}

The required size of $M_{1/2}$ is determined by the experimental value of the Higgs boson mass, $m_h^0 \sim 125$ GeV.
The contours of the Higgs boson mass and theoretical error $\Delta m_h$ are shown in Fig.~\ref{fig:mhiggs}. The Higgs boson mass is calculated by {\tt H3m} package~\cite{h3m} at the three loop level. The red (green) lines are drawn with $m_t=173.2$ GeV ($m_t=174.2$ GeV), taking into account a quite large uncertainty of the measured top pole mass as~\cite{top_tevatron, top_LHC} 
\bea
173.20 \pm 0.87 \ {\rm GeV}\,({\rm Tevatron}), \ \ 173.3 \pm 1.4 \ {\rm GeV}\,({\rm LHC}).
\eea
In addition, we estimated the theoretical error as $\Delta m_h=|m_{h}^{\rm 3loop}-m_{h}^{\rm 2loop}|/2$, that is the error of slowly convergent perturbative expansion.\footnote{This measure is suggested in Ref.~\cite{higgs_three}.  Here, we do not include the uncertainty of the top mass in $\Delta m_h$.} (The two-loop result $m_h^{\rm 2loop}$ is evaluated using {\tt FeynHiggs}~\cite{feynhiggs}.) This error is within a range of $0.5-3$ GeV. Here, we see that $M_2(=M_{1/2}) \gtrsim 3500$ GeV can be consistent with the experimental value of the Higgs boson mass~\cite{ATLAS_4l, ATLAS_2g, CMS_4l, CMS_2g}:
\bea
&&124.3\,\,^{+0.6}_{-0.5} \,({\rm stat.})\,^{+0.5}_{-0.3}\,({\rm syst.}){\rm \ GeV} \ \ ({\rm ATLAS} \ 4l),\nn
&&126.8\pm 0.2\,({\rm stat.}) \pm 0.7\,({\rm syst}.){\rm \ GeV}  \ \ ({\rm ATLAS} \ \gamma\gamma), \nn
&&125.8\pm 0.5\,({\rm stat.}) \pm 0.2\,({\rm syst.}){\rm \ GeV}  \ \ ({\rm CMS} \ 4l), \nn
&&125.4 \pm 0.5\,({\rm stat.}) \pm 0.6\,({\rm syst.}){\rm \ GeV} \ \ ({\rm CMS} \ \gamma\gamma).
\eea
 Notice that the Higgs boson mass is about 1\,GeV larger than the universal gaugino mass case for the fixed gluino mass, because the induced tri-linear coupling of stops is relatively large.

\begin{figure}[t]
\begin{center}
\includegraphics[scale=1.1]{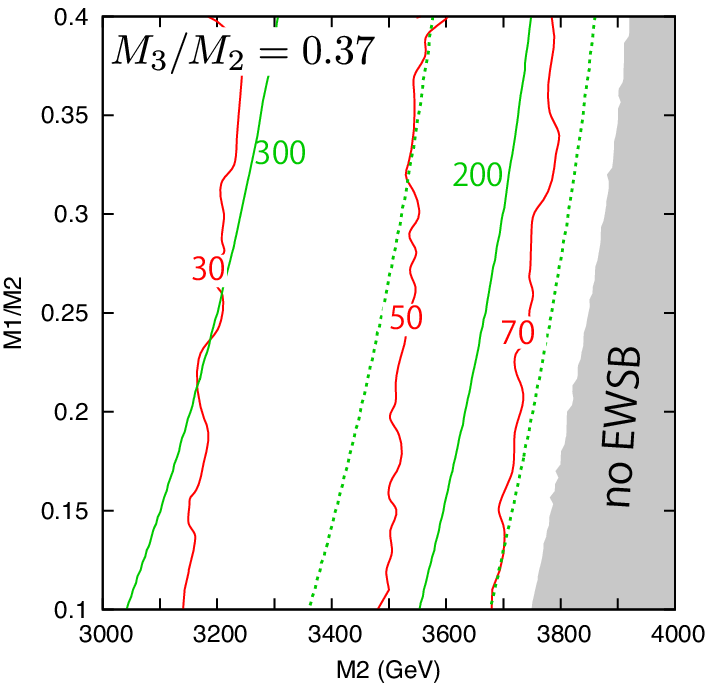}
\includegraphics[scale=1.1]{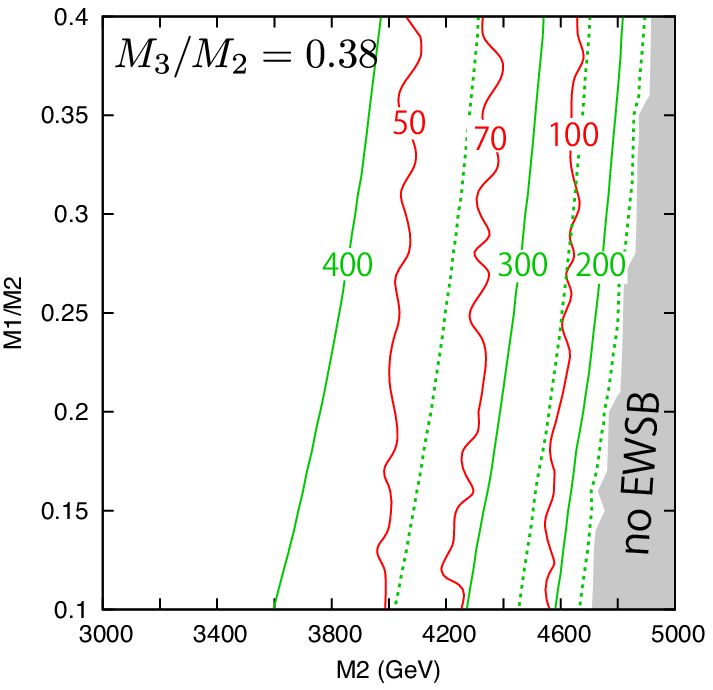}
\includegraphics[scale=1.1]{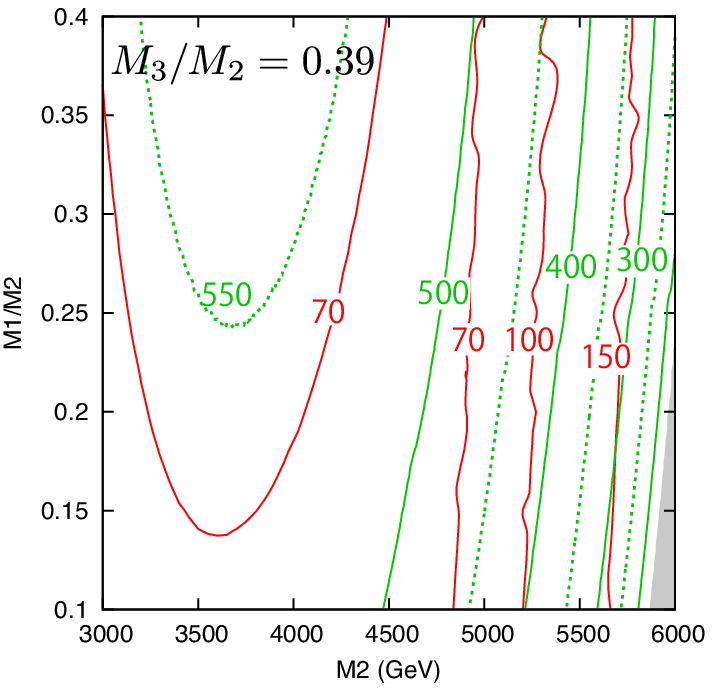}
\caption{The Higgsino mass parameter in the unit of GeV (green) and $\Delta$ (red). Here, $m_0=500$\,GeV and $\tan\beta=20$. In the gray shaded regions, the EWSB is unsuccessful.
}
\label{fig:ft_mu}
\end{center}
\end{figure}

Now, we consider the sensitivity of the EWSB scale by $M_{1/2}$. We take the fine-tuning measure as~\cite{finetuning}~\footnote{The fine-tuning with respect to the Higgs B-term is suppressed as $\sim (1/\tan^2\beta) \Delta_\mu$, which is negligible for large $\tan\beta$.} 
\bea
\Delta = \max (\Delta_{\mu^0}, \Delta_{M_{1/2}}), \ \ \Delta_a = \left| \frac{\partial \ln m_{\hat Z}}{\partial \ln a} \right|,
\eea 
where $\mu^0$ is the Higgsino mass parameter at the GUT scale. The sensitivity of $\mu^0$ is approximately given by
\bea
\Delta_{\mu^0} \sim \frac{2 \mu^2 }{m_Z^2}.
\eea
This is because $\mu$ is a SUSY invariant parameter and it is rather stable under radiative corrections. Notice that we can neglect $\Delta_{m_0}$ (see Eq.(\ref{eq:mhu2})). In Fig.~\ref{fig:ft_mu}, the contours of $\Delta$ and the Higgsino mass parameters are shown. The change of the bino mass parameter does not affect $\Delta$ so much. We find that the Higgsino mass is smaller than $500$ GeV for almost all region with $\Delta <100$, and hence, it can be discovered at the ILC.

%

\section{Mixed DM of Bino and Higgsino}
In our gaugino mediation model, the bino-like lightest neutralino can be a candidate for DM; The small $\Delta \lesssim 100$ requires the light Higgsino as shown in Fig.~\ref{fig:ft_mu}. Here, the wino is as heavy as gluino. On the other hand, the bino mass can be taken small keeping $\Delta \lesssim 100$ as long as $r_1 \sim 0.1$. 
There are two typical regions where the thermal relic abundance of the lightest neutralino $\chi_1^0$ can explain the observed value:  large enough Higgsino fraction or coannhilation with stau.

The relic abundance of the lightest neutralino is calculated using {\tt micrOMEGAs}~\cite{MOMG} and it is required to satisfy the experimental value~\cite{PDG}
\bea
\Omega_{\rm CDM}h^2 = 0.111 \pm 0.006 \ \ (1\sigma), \label{eq:relic}
\eea
where $h$ is the normalized Hubble parameter at the present universe. The contours of $\Delta$ and the blue strips where the relic abundance of the lightest neutralino is $\Omega_{\chi_1^0} h^2 \simeq 0.11$ are shown in Fig.~\ref{fig:SI} and \ref{fig:SD} for different $r_3$ and $m_0$. 
On the blue strips close to the regions where the stau is LSP, the stau-neutralino coannihilation is effective to reduce the abundance, other viable regions are simply due to the sizable Higgsino fraction.

In the case that $\chi_1^0$ has a large enough Higgsino fraction, the cross section between the neutralino and  nucleon tends to be large and a part of the parameter space is already excluded by the XENON100 experiment. When the abundance of the neutralino is reduced by the coannihilation mechanism with a stau, the fraction of the Higgsino is somewhat small. However, the mixing between the bino and Higgsino is still sizable and can be covered by XENON1T experiment~\cite{XENON1T}.

\subsection{XENON 100}
The region where the $\chi_1^0$ has a large Higgsino fraction is severely constrained by the direct detection experiment. The sizable mixing induces large contributions to the spin-independent (SI) cross section via Higgs boson mediated diagrams.
The most stringent constraint on the SI cross section is obtained from XENON100 experiment.  The upper bound of the SI cross section is (approximately) given by~\cite{Xenon100}
\bea
\sigma_{\rm SI} \lesssim 4.0 \times 10^{-45}\, {\rm cm}^2 \, \left(\frac{m_{\chi_1^0}}{200\, {\rm GeV}}\right),
\eea
for $m_{\chi_1^0} \gtrsim 100$\,GeV at 90\,\% confidential level. Here, $m_{\chi_1^0}$ is the mass of the lightest neutralino.

\begin{figure}[t]
\begin{center}
\vspace{-120pt}
\includegraphics[scale=1.05]{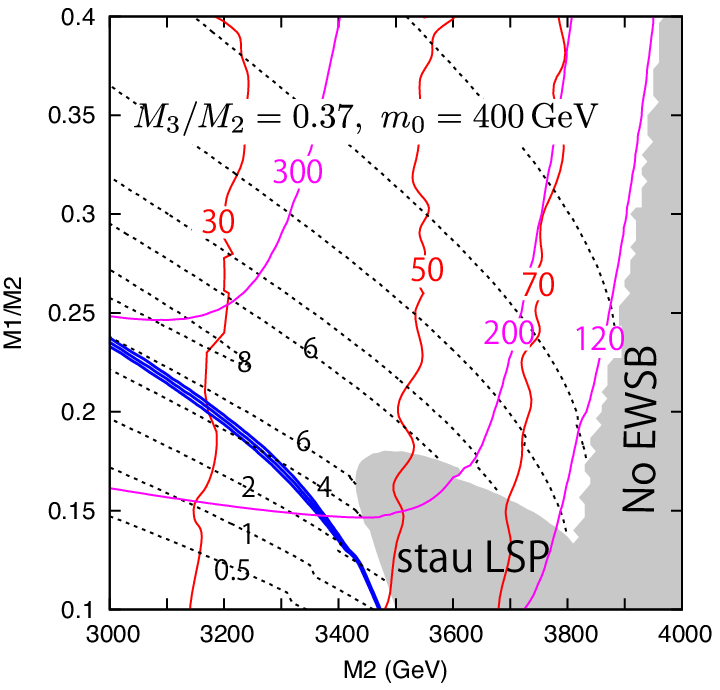}
\includegraphics[scale=1.05]{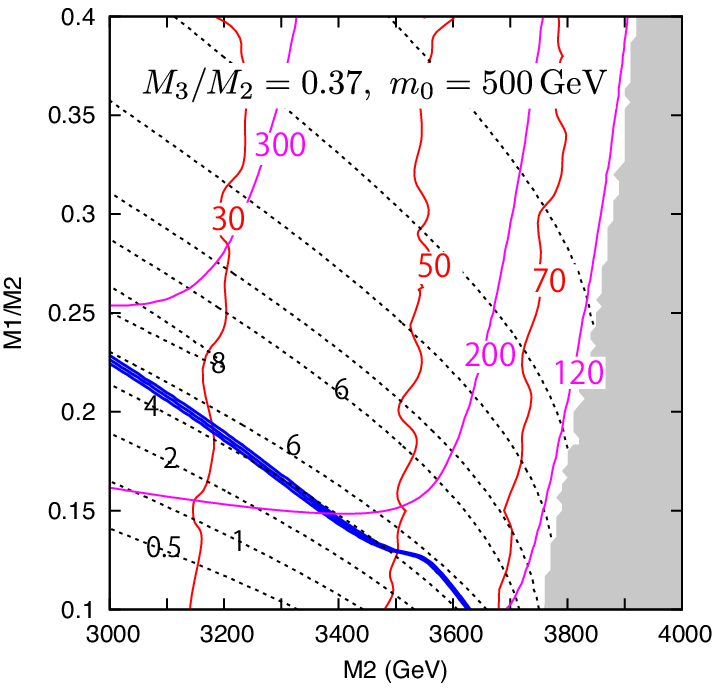}
\includegraphics[scale=1.05]{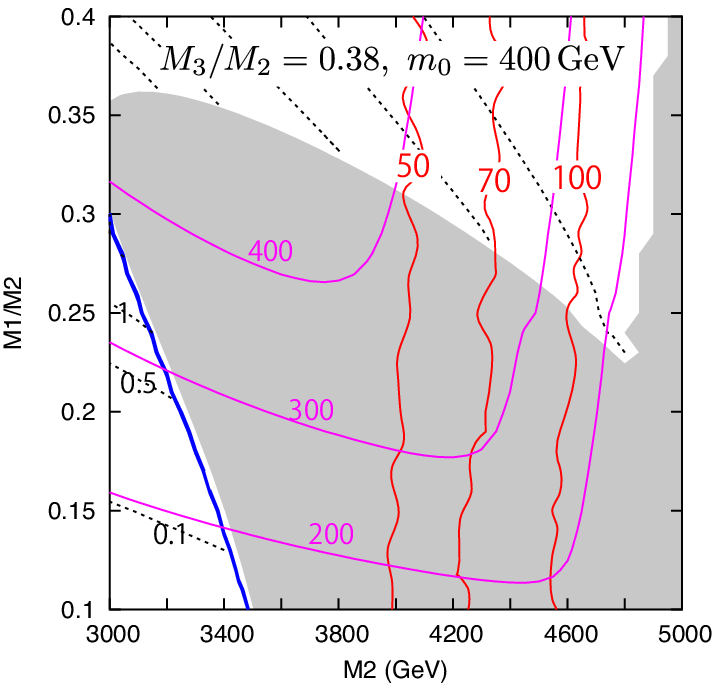}
\includegraphics[scale=1.05]{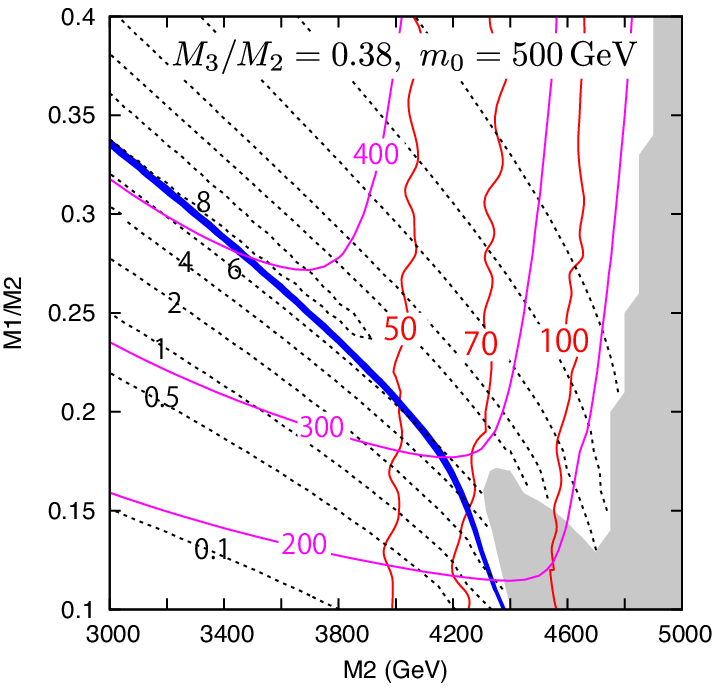}
\includegraphics[scale=1.05]{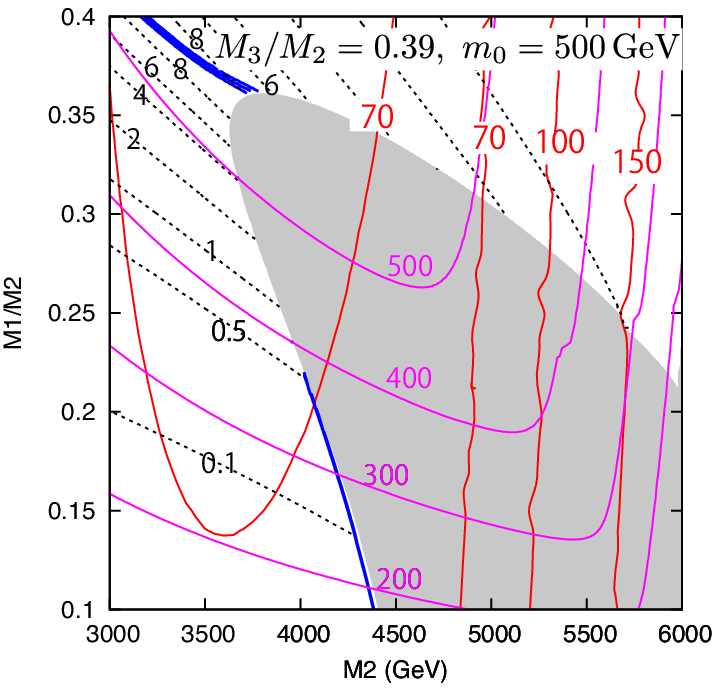}
\includegraphics[scale=1.05]{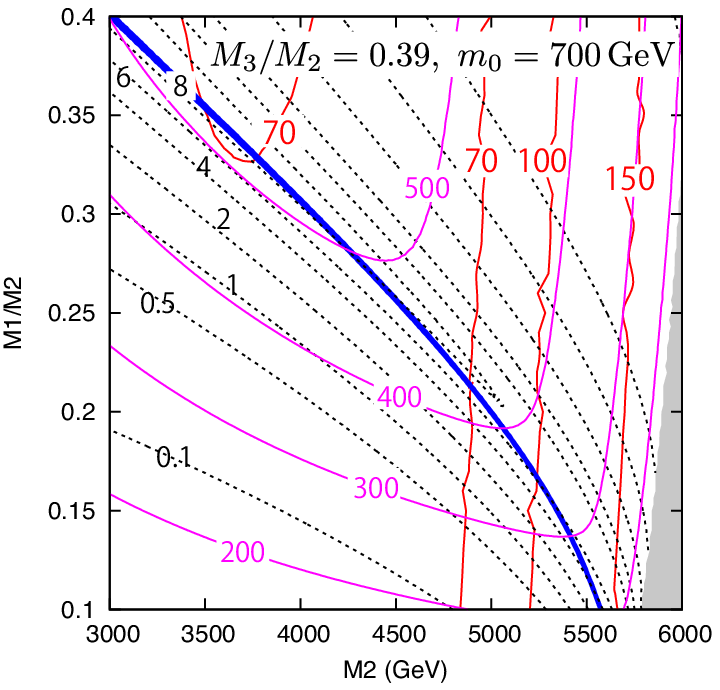}
\caption{SI cross section in the unit of $10^{-45}$ cm$^2$. The neutralino mass is shown as solid magenta line. The blue solid line corresponds to $\Omega_{\chi_1^0} h^2 \simeq 0.11$. In the gray shaded regions, the EWSB is unsuccessful or the stau is the LSP.}
\label{fig:SI}
\end{center}
\end{figure}

In Fig.~\ref{fig:SI}, we show the contours of the SI cross sections in the unit of $10^{-45}$ cm$^2$. In the calculation, we use {\tt micrOMEGAs} package and the strange quark content of the neucleon is taken as $f_s=0.009$ which is the result of the recent lattice calculation~\cite{lattice_fs}.  With $f_s\simeq 0.26$ (default value of the {\tt micrOMEGAs}), the SI cross section becomes about twice as that with $f_s = 0.009$. In the regions where the masses of the stau and neutralino are not degenerated, the SI cross section is marginally consistent with the current experimental bound from the XENON100. In the region where the coannihilation takes place, the cross section is somewhat smaller and the current bound can be avoided easily, but the cross section is still large due to the non-negligible mixing between the bino and Higgsino. At the XENON1T experiment, the sensitivity is expected to improve two order of magnitude with 2 years operation~\cite{XENON1T}, and the whole regions consistent with the observed dark matter abundance will be covered. Note that in the region where the mass difference is smaller than the tau mass, the life-time of the stau is long and can be detected at the LHC using charged track~\cite{CHT}. 

The XENON100 experimental data also constrains the spin-dependent (SD) cross section~\cite{SD_XENON}. However, more stringent constraint comes from IceCube experiment.

\subsection{IceCube}
The high energy neutrino flux is induced by the neutralino anihilation in the Sun. This neutrino can be detected by IceCube experiment, and due to the absence of positive signals, the size of the neutralino-necleon scattering cross section determining the capture rate in the Sun is bounded from above.
The current bound of the SD cross section between the proton and neutralino is given by~\cite{IceCube}
\bea
\sigma_{\rm SD} \lesssim 10^{-40}\, {\rm cm}^2\ {\rm for}\  m_{\chi_1^0} =(100-500)\, {\rm GeV}, \label{eq:icq}
\eea
where the neutralinos annihilate into $W^+ W^-$, exclusively. In our case, the neutralinos dominantly (but not exclusively) annihilate into top pairs 
apart from the coannihilation region, and the constraint is close to Eq.~(\ref{eq:icq})~\cite{ICC_tt}.

\begin{figure}[t]
\begin{center}
\vspace{-120pt}
\includegraphics[scale=1.05]{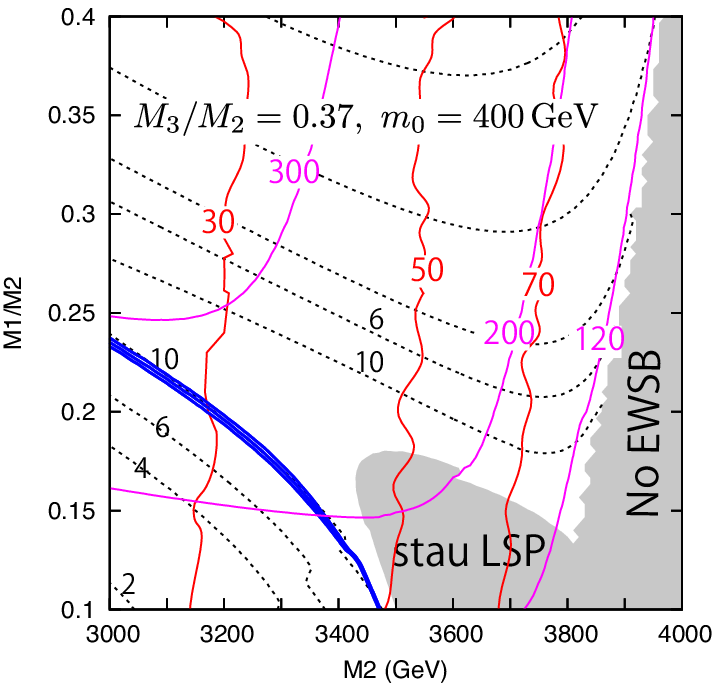}
\includegraphics[scale=1.05]{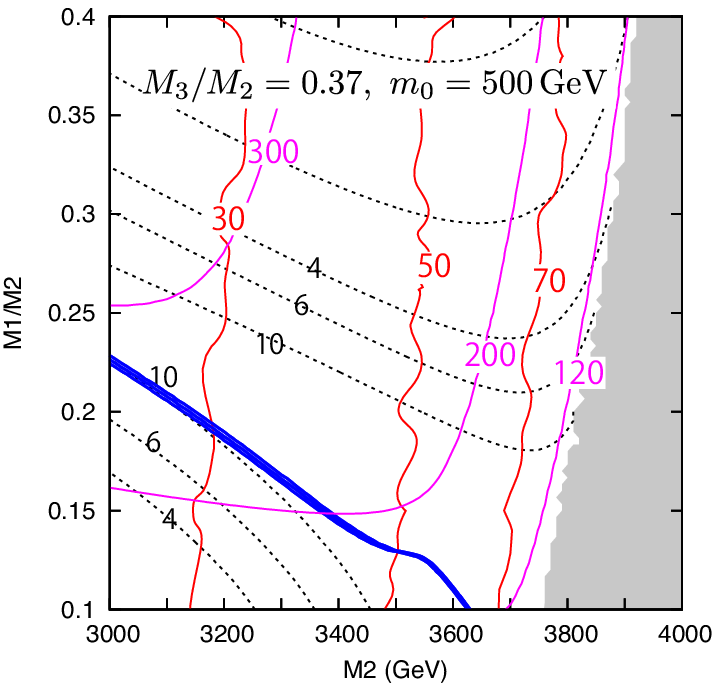}
\includegraphics[scale=1.05]{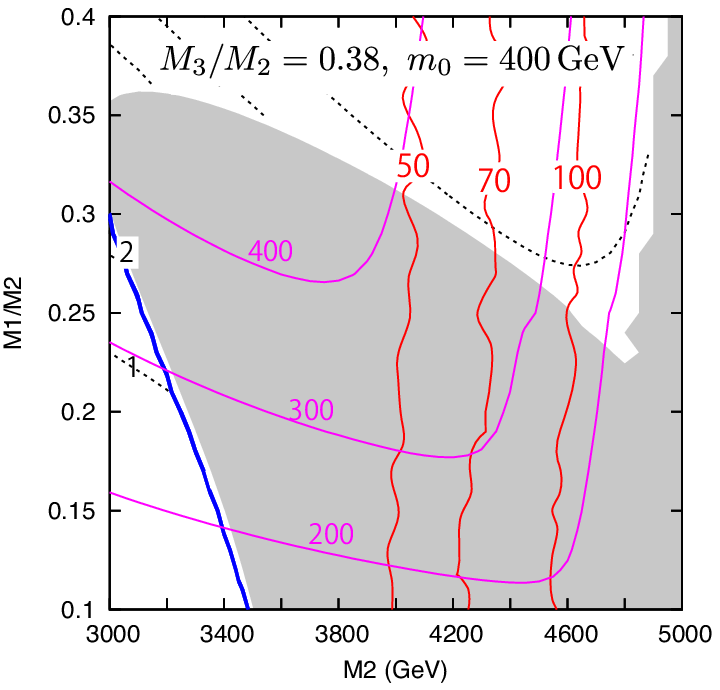}
\includegraphics[scale=1.05]{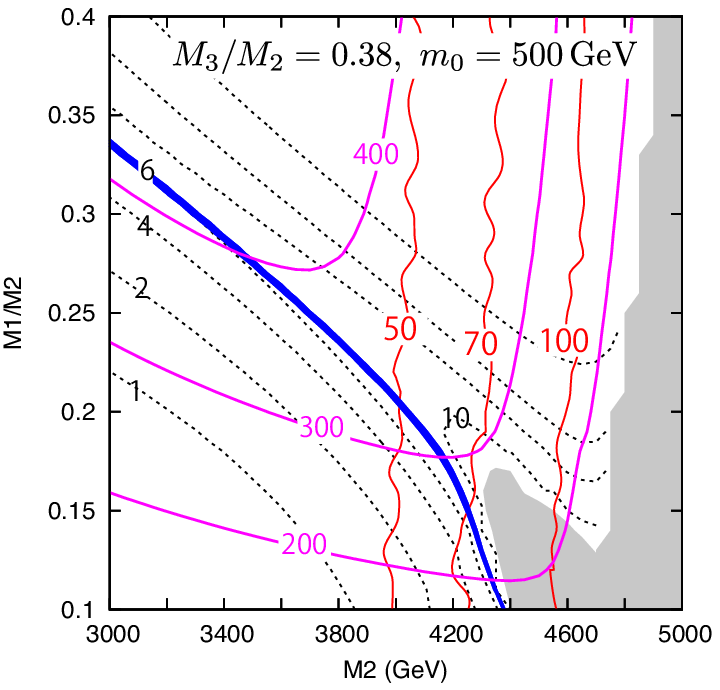}
\includegraphics[scale=1.05]{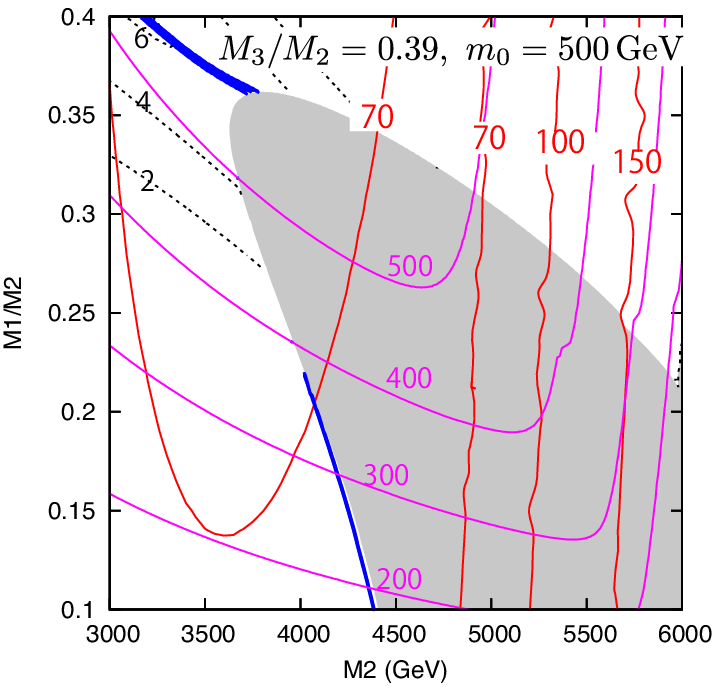}
\includegraphics[scale=1.05]{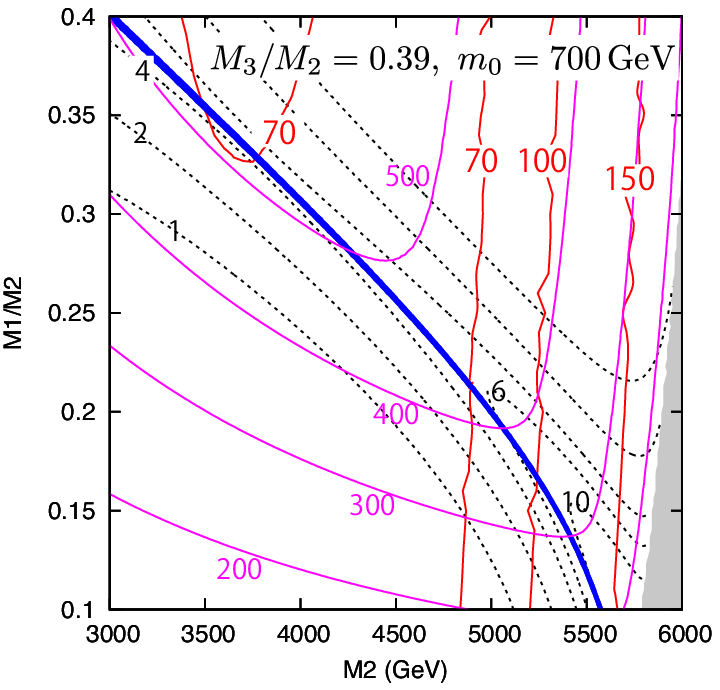}
\caption{SD cross section in the unit of $10^{-41}$ cm$^2$. The neutralino mass is shown as solid magenta line.}
\label{fig:SD}
\end{center}
\end{figure}

In Fig.~{\ref{fig:SD}}, the contours of the SD cross section are shown in the unit of $10^{-41}$ cm$^2$. In the case of $M_1/M_2=0.37$, the region with small $\Delta \sim 30$ is likely to be close to the current exclusion bound. Such a very low $\Delta$ region is expected to be covered in near future. In the case $M_1/M_2=0.38$ and $M_1/M_2=0.39$, the constraint can be avoided easily in most parameter space consistent with $\Omega_{\chi_1^0} h^2 \simeq 0.11$.

\section{Conclusions and discussion}

The focus point gaugino mediation model with $\mathcal{O}(100)\, {\rm GeV}$ gravitino is consistent with the observed Higgs boson mass with the mild fine-tuning. We have shown that the thermally produced bino-like neutralino can explain the present dark matter density. The scattering cross section between the neutralino and neucleon is rather large as a consequence of the sizable Higgsino fraction, and hence, some of the region is already excluded by the XENON100 experiment. The other regions are also expected to be tested at the future dark matter experiments such as XENON1T and IceCube.

Finally, let us discuss the ratio of the gluino mass to wino mass, $M_3/M_2=0.37-0.38 \sim 3/8$, which is crucial for the mild fine-tuning of $\Delta \lesssim100$. As shown below, this gaugino mass ratio can be obtained as a result of more fundamental physics. 
Suppose that there exists a non-anomalous discrete R-symmetry $Z_{NR}$ in the more fundamental theory and the present model is its low-energy effective theory. 
The mixed gauge anomalies $Z_{NR}$--$SU(2)$--$SU(2)$ and $Z_{NR}$--$SU(3)$--$SU(3)$ in the present model are given by~\cite{znr_anomaly, znr_anomaly2}
\bea
{\bf A}_2 &=& 4 + n_g\, ( 3 \cdot r_{\bf 10} + r_{\bar{\bf 5}}-4)  + (r_u + r_d -2), \nn
{\bf A}_3 &=& 6 + n_g\, ( 3 \cdot r_{\bf 10} + r_{\bar{\bf 5}}-4), 
\eea
where $r_u (r_d)$ is $Z_{NR}$ charge of $H_u (H_d)$. The charges of $(Q, \bar{U}, \bar{E})$ are denoted by $r_{\bf 10}$ and those of $(\bar{D}, L)$ are $r_{\bar{\bf 5}}$. The number of the generations is set to be  $n_g=3$. The Yukawa terms in the superpotential should be consistent with $Z_{NR}$, gives additional constraints as~\footnote{The seesaw mechanism can be consistent with $Z_{NR}$. The required conditions are $r_1 + r_u + r_{\bar{\bf 5}}=2$ and $2 r_1=2$, where $r_1$ is the $Z_{NR}$ charge of the right-handed neutrino superfield.}
\bea
r_u + 2 r_{\bf 10} =2 \ \ {\rm mod}\,\, N,  \ \ r_d + r_{\bf 10} + r_{\bar{\bf 5}}=2 \ \ {\rm mod}\,\, N.  
\eea
By using above conditions, ${\bf A}_2$ and ${\bf A}_3$ become
\bea
{\bf A}_2 &=& 4 - n_g\, (r_u +r_d)  + (r_u + r_d -2) \ \ {\rm mod}\,\, N, \nn
{\bf A}_3 &=& 6 - n_g\, (r_u+r_d)  \ \ {\rm mod}\,\, N.
\eea
Requiring that $\mu$ term is generated by Giudice Masiero mechanism~\cite{giudice_masiero} or R-breaking mechanism~\cite{rbm}, we get $r_u + r_d=0\, {\rm\  mod\  }N$ (and $r_u + r_d \neq 2$). As a result, mixed gauge anomalies become
\bea
{\bf A}_2=2\  {\rm mod}\,\, N, \ \ {\bf A}_3 = 6 \  {\rm mod}\,\, N.
 \eea
 
We suppose that these anomalies are cancelled by a shift of the imaginary part of the Polonyi field $Z$. The relevant terms are given by
\bea
\frac{k_2}{32\pi^2} \int d^2\theta \frac{Z}{M_*}( W^a_\alpha)_2 (W^{a\,\alpha})_2,\ \ \frac{k_3}{32\pi^2} \int d^2\theta \frac{Z}{M_*} (W^a_\alpha)_3 (W^{a\,\alpha})_3, 
\eea
where $(W_\alpha^a)_2$ and $(W_\alpha^a)_3$ consist of $SU(2)$ and $SU(3)$ vector multiplets, respectively. The mass scale $M_*$ is determined by the fundamental theory (e.g., $M_*\sim 10^{16}$ GeV).
Here, we assume that ${\rm Im}(Z/M_*)$ transforms as ${\rm Im}(Z/M_*) \to {\rm Im}(Z/M_*) + (2\pi l' /N)$ under $Z_{NR}$, while a fermionic component of a chiral super field transforms as $\psi_i \to \psi_i \exp\,[i (r_i-1) (2\pi l'/N)]$ ($l'$ is an integer and $r_i$ is a $Z_{NR}$ charge with $i={\bf 10}, {\bf \bar{5}}, {H_u}, {H_d}$). Then, the conditions for anomaly cancellation are written as
\bea
2  + k_2 =0\  {\rm mod}\,\, N, \ \  6 + k_3=0 \  {\rm mod}\,\, N.
\eea

Let us now assume even number of $N$, since otherwise the R-parity is broken by the constant term of the superpotential~\cite{STY} and see if the ratio of the gluino mass to wino mass $M_2/M_3=k_2/k_3=8/3$ can be obtained. In the minimal case $Z_{4R}$, solutions which satisfy the condition $k_2/k_3=8/3$ are given by 
\bea
k_2=2+4n=8l, \ k_3=2+4m=3l,
\eea
where $n$, $m$ and $l$ are integers. Because of the condition for $k_2$, there is no solution in $Z_{4R}$. The next minimal case is $Z_{6R}$ and solutions in this case are given by
\bea
k_2=4+6n=8l, \ k_3=6m=3l.
\eea
We see a solution $n=2$, $m=1$ gives the desirable gaugino mass ratio $M_2/M_3=k_2/k_3=8/3$.  
It is very surprising that the minimal and non-trivial solution ($m \neq 0$) is found.
These non-universal gaugino masses may be consistent with the product group unification~\cite{PGU}.
 It might be interesting to know that the desirable ratio can be obtained also when the wino mass and gluino mass are simply proportional to $M_{1/2}/{\rm dim}(G_i)$. Here, ${\rm dim}(G_2)$ (${\rm dim}(G_3)$) is a dimension of an adjoint representation of $SU(2)$ ($SU(3)$). We will discuss such more fundamental theories in a future publication.

\section*{Acknowledgements}
The work of NY is supported in part by JSPS Research Fellowships for Young Scientists.
This work is also supported by the World Premier International Research Center Initiative (WPI Initiative), MEXT, Japan.


\end{document}